# ACCURACY OF AN ATOMIC MICROWAVE POWER STANDARD

David C. Paulusse, Nelson L. Rowell and Alain Michaud

Institute for National Measurement Standards, National Research Council of Canada
M-36, 1200 Montreal Road, Ottawa, ON, Canada, K1A-0R6, E-mail: Alain.Michaud@nrc-cnrc.gc.ca

**Abstract**

We have studied the accuracy of the atomic microwave power standard. The atoms are cooled and kept in a Magneto-Optical Trap (MOT), then dropped through a terminated transmission line (a rectangular, R-70 type, waveguide). The measurement of the internal atomic state allows an accurate determination of the transmitted microwave power.

**Introduction**

The introduction of laser cooling has been a revolution in metrology and atomic physics in general. The controlled variation of atomic properties using applied electromagnetic fields is useful in a wide range of precise measurements. Using similar techniques, microwave power measurements could also be improved, especially since the accuracy of the actual standards is limited [1].

The initial experiment has shown that a linearity of 0.3 % over a 20 dB range could be achieved [2,3]. Also, the evaluation of the of the power inside the resonant cavity of an atomic fountain was shown to be in agreement with direct measurement to a level of about five percent [4].

The present experiment will allow the accurate measurement of the power of microwave radiation transmitted into a rectangular waveguide. The following is a short description of the apparatus.

**Description of the Apparatus**

The setup shown in Fig. 1 is similar to that of our previous experiment and uses the same time sequence. The Rubidium atoms are captured in a standard magneto-optical trap (MOT). Simply shutting off the lasers and letting the atoms fall into the interaction region does the measurement. If necessary, the interaction time can be modified by changing the height above the waveguide or by launching the atoms into the waveguide.

The transmission line (Fig. 2) is coupled to the output of a transfer standard on one side and to a termination or a matched detector on the other side. The system can then be used to calibrate either the leveled source or the matched detector. [1]

The microwave generator should be pulsed in order to evaluate the field distribution as the atoms cross the waveguide. In this way, the filling factor and interaction time can be determined. The population inversion is then measured as a function of the field amplitude, and the knowledge of the interaction time (assuming that the detector is linear), allows an accurate determination of the calibration factor.

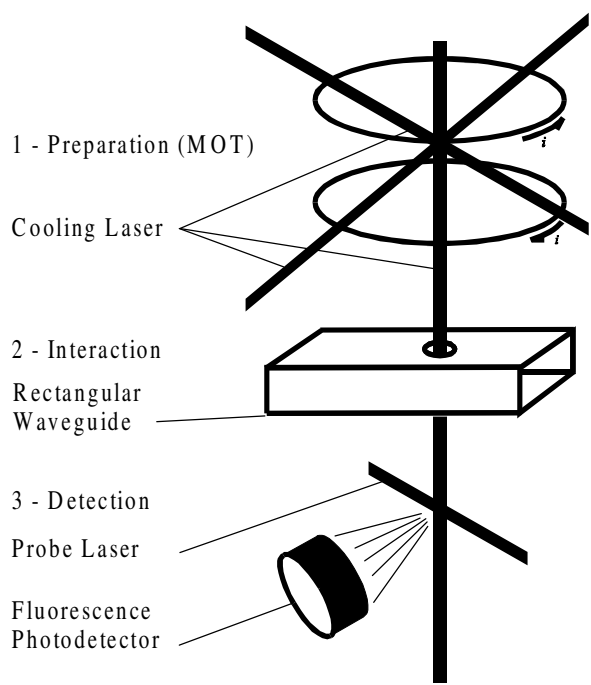

Figure 1. Schematic of the experimental system.

The transmission line (Fig. 2) has to be characterized in order to evaluate the accuracy of the standard. It is equipped with access holes installed on the broad walls, which were optimized to minimize any leakage of the radiation outside of the guide.

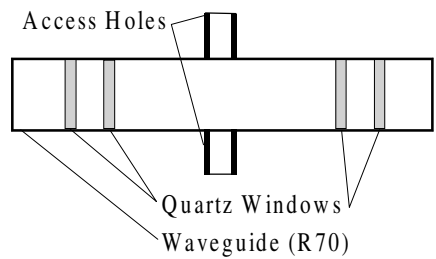

Figure 2. Diagram of the transmission line. The device is symmetrical and reversible.

The most critical components include the two hermetic windows, necessary in order to be compatible with standard "in air" instruments. They should be reflectionless at the frequency of operation. Each window consists of two identical thin quartz plates. This design offers an efficiency advantage over the traditional design.

Figure 3 shows a plot of the insertion loss of the complete line as a function of the frequency. Practically all the residual loss, in total about one percent, is due the normal attenuation of the metallic waveguide.

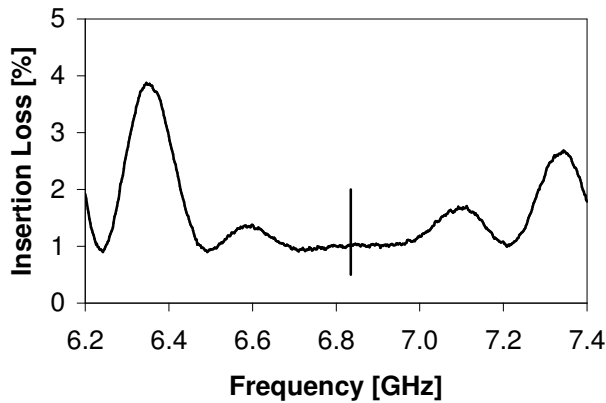

Figure 3. Frequency response of the transmission line shown in Fig. 2. The vertical line shows the frequency of operation (6.8347 GHz).

### Outline of the presentation

In this paper we report on the progress of this work. We show figures describing the experiment and we present our latest experimental results. We also discuss the various sources of errors in the accuracy of this system.

### References

[1] A. Fantom, *Radio Frequency and Microwave Power Measurement*, London, Peter Peregrinus Ltd., 278 pp., 1990. (IEE press: http://www.iee.org/)

[2] D.C. Paulusse, N.L. Rowell, A. Michaud, "Realization of an Atomic Power Standard", *2002 Conference on Precision Electromagnetic Measurements,* pp. 194-195, Ottawa, 16-21 June 2002.

[3] D.C. Paulusse, N.L. Rowell, A. Michaud, "Microwave Power Standard using Cold Atoms", *Eighteenth International Conference on Atomic Physics (ICAP 2002), Poster Presentation Abstracts,* p. 329, Cambridge, July 28 – August 2, 2002. Online:
http://www.wspc.com.sg/icap2002/article/2423055.pdf

[4] E.A. Donley, T.P. Crowley, T.P. Heavner, B.F. Riddle "Quantum-Based Microwave Power Measurement Performed with a Miniature Atomic Fountain", *Proceedings: 2003 IEEE International frequency Control Symposium,* p. 135-137, Tampa Bay, May 4-8, 2003.